\documentstyle[titlepage,11pt]{article}
\title{\bf Kinetic roughening of surfaces: Derivation, solution and 
application of linear growth equations}

\author{S. Majaniemi$^{1}$, T. Ala--Nissila$^{1-3}$ and
J. Krug$^{4}$
\\ \\
$^1$Research Institute for Theoretical Physics \\
P.O. Box 9 (Siltavuorenpenger 20 C) \\
FIN--00014 University of Helsinki, Finland \\
\\
$^2$Department of Physics \\
P.O. Box 692 \\
Tampere University of Technology \\
FIN--33101 Tampere, Finland \\ \\
$^3$Department of Physics \\
Box 1843 \\
Brown University \\
Providence R. I. 02912, U.S.A. \\ \\
$^{4}$Institut f\"ur Festk\"orperforschung \\
Forschungszentrum J\"ulich \\
D-52425 J\"ulich, Germany \\
\\
$^{5}$Present address: \\
Fachbereich Physik, Universit\"at GH Essen \\
D-45117 Essen, Germany \\
}

\date{January 16, 1996}

\begin{document}

\maketitle

%%%%%%%%%%%%%%%%%%%%%%%%%
% 	ABSTRACT        %
%%%%%%%%%%%%%%%%%%%%%%%%%
\begin{abstract}
We present a comprehensive analysis of a linear growth model, which combines
the characteristic features of the Edwards--Wilkinson and noisy Mullins
equations. This model can be derived from microscopics and it describes the
relaxation and growth of surfaces under conditions where the nonlinearities
can be neglected. We calculate in detail the surface width and various
correlation functions characterizing the model. In particular, we study the
crossover scaling of these functions between the two limits described by the
combined equation. Also, we study the effect of colored and conserved noise
on the growth exponents, and the effect of different initial conditions. The
contribution of a rough substrate to the surface width is shown to decay
universally as $w_i(0) (\xi_s/\xi(t))^{d/2}$, where $\xi(t) \sim t^{1/z}$ is
the time--dependent correlation length associated with the growth process,
$w_i(0)$ is the initial roughness and $\xi_s$ the correlation length of the
substrate roughness, and $d$ is the surface dimensionality. As a second
application, we compute the large distance asymptotics of the height
correlation function and show that it differs qualitatively from the
functional forms commonly used in the intepretation of scattering 
experiments.
\\
\noindent
PACS numbers: 68.55.-a, 68.10.Jy, 05.40.+j.
\end{abstract}

\textheight 21cm
\textwidth 14.5cm
\oddsidemargin 0.96cm
\evensidemargin 0.96cm
\topmargin -0.31cm
\raggedbottom
\parindent=0.0cm
\newcommand{\beq}{\begin{equation}}
\newcommand{\eeq}{\end{equation}}
\section{Introduction}

The dynamics of interfaces ranging from
dendritic growth
%\cite{Lan78}
to flame front propagation
%\cite{Sne92}
can often be
described by relatively simple evolution equations \cite{Pel88}. Typically
such
evolution equations are given in terms of partial differential equations
with a stochastic noise component. Perhaps the best known example is the
nonlinear Kardar--Parisi--Zhang (KPZ) equation \cite{KPZ}
which describes kinetic roughening of randomly driven interfaces such as
growing surfaces or flame fronts in forest fires \cite{Pro95}.
Complete understanding of these
nonlinear equations is still mostly lacking.

Under certain circumstances
discussed below the relevant nonlinearities may be so weak that a fully
{\it linear} model can provide an adequate description. One example
is the growth model of Edwards and Wilkinson
(EW) \cite{Edw82}, which describes the sedimentation of
granular particles under gravitation. Another important model is the
noisy Mullins equation discussed by Wolf and Villain and others
\cite{Wol90,DT,Golub}
(the Mullins--Wolf--Villain (MWV) equation)
in the context of molecular beam epitaxy (MBE). Being
linear, both of these equations have been analyzed in some detail.
However, recently it has been shown both from macroscopic
arguments \cite{Vil91} and through more microscopic
derivations
\cite{Zangwill1,Zangwill2}
that for some cases involving surface
diffusion and desorption, a more general linear equation of the form
\beq
\partial_t h = \nu_1\nabla^2 h +\nu_2\nabla^4 h + \eta
\label{combi}
\eeq
emerges, where $h=h(\vec x,t)$
is the surface height above a $d$--dimensional substrate, $\eta(\vec x,t)$ is
a noise
term, and $\nu_1$ and $\nu_2$ are parameters. Since
the gradient terms of Eq. (\ref{combi}) are simply a combination of the
EW and MWV equations, we call it the
{\it Combined Linear Growth} (CLG) equation. Stability requires that
$\nu_1 \geq 0$ and $\nu_2 \leq 0$. While it is physically possible that
$\nu_1 < 0$ (see Sec. 2.2), the treatment of this case requires the
inclusion of additional nonlinear terms in (\ref{combi}), and will not
be addressed here.

The purpose of the current work is to present a detailed
analysis of Eq. (1), which is missing so far.
This is useful for two main reasons. First, the calculations
in this work generalize the previous results obtained for
the EW and MWV equations which are somewhat incomplete and
scattered
in the literature
\cite{Edw82,Wol90,Golub,Vil91,Fam86,Lam91,Nat92,Ama93,Sar94,thh}.
Second, Eq. (\ref{combi}) is the simplest example of a growth equation
with an {\em intrinsic length scale}. Balancing the two gradient
terms in (\ref{combi}) one finds that they become comparable at the scale
\beq
\ell^\ast = \sqrt{|\nu_2|/\nu_1}.
\label{lstar}
\eeq
The kinetic roughening process is governed by the fourth derivative term
on scales smaller than $\ell^\ast$ but the second order term dominates on
scales larger than $\ell^\ast$. Physically, the two terms represent different
relaxation mechanisms, through surface diffusion (fourth derivative),
evaporation--condensation or step edge barriers (second derivative)
\cite{Vil91,mullins}.
Of course, writing a continuum equation with an intrinsic scale is meaningful
only if this scale much exceeds the microscopic cutoff, given by
the lattice spacing $a$; the detailed estimated of $\nu_1$ and $\nu_2$ derived
in Sec. \ref{derivation of linear growth model} show that
$\ell^\ast$ is indeed
{\em mesoscopic}, in the sense that $\ell^\ast \gg a$, under typical
conditions.
Mesoscopic intrinsic length scales such as domain \cite{Pal94} or terrace
\cite{Vil92}
sizes play an important role in the kinetic roughening of real surfaces.
Moreover, the competition and interplay between different relaxation and
roughening mechanisms
is probably typical in many experimental situations; for this reason combined
linear
equations like (\ref{combi}) have already been used extensively in the
analysis of
experimental data \cite{tong94}.

A major advantage of working with {\em linear} growth equations is that they
allow us to explicitly compute any statistical quantity of interest, rather
than just extracting the values of scaling exponents, which have been the
focus of most previous studies of kinetic roughening \cite{reviews}. We
will exploit this fact to address two questions
of direct experimental relevance: The evolution of the initial
substrate roughness during growth, and the shape of the height--height
correlation function. In both cases we find that the heuristic expressions
commonly employed in the experimental literature are
{\em incompatible} with the explicit calculations.

The organization of the paper is as follows. We shall first briefly review
the physical background and various arguments leading to Eq.
(\ref{combi}), and in particular various
interpretations of the coefficients $\nu_1$ and $\nu_2$. Following this,
Sec. \ref{solutions for various quantities} contains a full solution of
the CLG model in terms of the relevant measures of the surface roughness.
We concentrate on the scaling behavior and finite size dependence of the
surface width $w(L,t)$, the equal time height--height correlation function
$G(x,t)$, and the saturated height--height correlation function $C_s(x,t)$.
In particular, we study in detail the {\it crossover behavior}
of these quantities to the well known limits given by the EW and
MWV equations. In Sec. 4 we discuss the experimentally important effects
of substrate roughness, the shape of the height correlation function and
the influence of long ranged noise correlations. Finally, summary and
conclusions are given in Sec. 5.
\section{Derivation of the Linear Growth Model}
\label{derivation of linear growth model}
In this section we provide some microscopic justification for the
combined linear growth equation (\ref{combi}). We consider two different
physical situations, corresponding to a surface in thermal equilibrium
(Sec. \ref{Equilibrium}) and a vicinal surface growing in the
step--flow mode (Sec. \ref{Schwoebel}), respectively.
In both cases the fourth order derivative
term in (\ref{combi}) reflects capillarity--driven surface diffusion
\cite{mullins}, while the second order term will be seen to arise from
distinct mechanisms.

\subsection{Equilibrium Dynamics of a Solid--on--Solid Model}
\label{Equilibrium}

We consider a one--dimensional solid--on--solid surface described by a set of
integer height variables $h_i$ defined on a lattice. The energy of
the surface is given by the Hamiltonian \cite{weeks}
\beq
\label{SOS}
{\cal H} = (K/2) \sum_i \vert h_i - h_{i+1} \vert.
\eeq
The surface evolves according to the following dynamic processes
\cite{Zangwill1}: Particles are deposited ($h_i \to h_i + 1$) at a constant
rate $F$; they evaporate ($h_i \to h_i -1$) at rate
\beq
\label{evap}
W_i^{\rm ev} = k_0 \exp[-(E'_S + n_i E'_N)/k_B T]
\eeq
and jump from site $i$ to site $j = i \pm 1$ at rate
\beq
\label{jump}
W_{ij}^{\rm diff} = k_0 \exp[-(E_S + n_i E_N)/k_B T].
\eeq
Here $k_0$ defines some microscopic hopping rate of the order of a typical
phonon frequency. We will generally measure time in units of $k_0^{-1}$, so
that effectively $k_0 = 1$; likewise the basic length unit will be
provided by
the lattice constant. The energy barriers in (\ref{evap}) and (\ref{jump})
each have a substrate contribution ($E'_S$ and $E_S$) and a bonding
contribution proportional to the lateral coordination number
\beq
\label{ni}
n_i = \theta(h_{i+1} - h_i) + \theta(h_{i-1} - h_i)
\eeq
which takes the values $n_i = 0,1,2$ in one dimension. Detailed balance
relative to (\ref{SOS}) holds if
\beq
\label{detbal}
E_N = E'_N = K \;\;\; {\rm and} \;\;\; F = k_0 \exp[-(E'_S + K)/k_B T].
\eeq
The distinguishing feature of the diffusion rates (\ref{jump}) is that they
depend only on the environment at the initial site $i$. It was shown elsewhere
\cite{Kru95} how this fact can be used to {\em exactly} derive the continuum
equation of motion for the surface, in the case where only surface diffusion
is allowed. Here we generalize the approach of \cite{Kru95} to include
desorption and deposition. Note, however, that this derivation is only valid
when the surface is in {\em equilibrium} with the vapor, as expressed by the
second condition in (\ref{detbal}).

{}From the master equation of the process one easily derives the following
equation of motion for the ensemble averaged height \cite{Zangwill1,Kru95},
\beq
\label{averageh}
\frac{d}{dt} \langle h_i \rangle = \frac{1}{2} e^{- E_S/k_B T} (\nabla^2
\lambda)_i - e^{- E'_S/k_B T} \lambda_i + F
\eeq
where $(\nabla^2 \lambda)_i = \lambda_{i+1} + \lambda_{i-1} - 2 \lambda_i$
denotes the lattice Laplacian, and  $\lambda_i \equiv \langle \exp[- (K/k_B T)
n_i]\rangle$. This quantity is related to the local chemical potential $\mu_i$
\cite{Kru95},
\beq
\label{lambdamu}
\lambda_i = \exp[-(K - \mu_i)/k_B T].
\eeq
We now pass to the continuum limit $\langle h_i \rangle, \mu_i \to h(x,t),
\mu(x,t)$, where $h(x,t)$ and $\mu(x,t)$ are averages taken over some large
region (still small on the macroscopic scale) centered around $i = x$.
The local chemical potential $\mu(x,t)$ is then determined by the local
surface curvature via a Gibbs--Thomson relation
\beq
\label{Gibbs}
\mu = - \hat \gamma(\nabla h) \nabla^2 h
\eeq
where the stiffness $\hat \gamma$ is a nonlinear function of the local
surface slope that can be directly computed from the Hamiltonian
(\ref{SOS}) \cite{Kru95}; at zero tilt ($\nabla h =0$)
\beq
\label{stiffness}
\hat \gamma(0) = k_B T [\cosh(K/2 k_B T) - 1]\,\, .
\eeq
For slowly varying, macroscopic profiles the typical curvatures are
small, so that (\ref{lambdamu}) can be expanded in $\mu$. This results in the
macroscopic equation
\beq
\label{macro1}
\partial_t h  =  - \frac{1}{2}  (k_B T)^{-1} e^{-(E_S + K)/k_B T} \nabla^2
[\hat\gamma (\nabla h) \nabla^2 h] + 
\eeq
$$
+ (k_B T)^{-1} e^{-(E'_S + K)/k_B T} \hat \gamma (\nabla h)\nabla^2 h
$$
which can be used e.g. to predict the decay of periodic surface modulations
\cite{Kru95}. Note that the deposition term has disappeared due to the second
of the detailed balance conditions (\ref{detbal}).
Because of the orientation dependence of the stiffness, (\ref{macro1})
is highly nonlinear.

Here we are primarily interested in the mesoscopic fluctuations around a
surface that is on average flat. Eq. (\ref{macro1}) can then be linearized by
expanding $\hat \gamma$ around the average orientation $u = \langle \nabla h
\rangle$. It is not strictly speaking consistent to keep only the terms linear
in $h$, since the nonlinearities arising from the expansion of the stiffness
in the second term of (\ref{macro1}) (such as $(\nabla h)^2 \nabla^2 h$) are
more relevant in the renormalization group sense than the linear fourth order
term; however we will ignore this difficulty in the interest of obtaining an
analytically tractable model. Thus we arrive at the two systematic terms in
(\ref{combi}), and identify the coefficients as
\beq
\label{nu1nu2}
\nu_1 =  (k_B T)^{-1}  \hat \gamma (u) e^{-(E'_S + K)/k_B T} , \;\;\;
\nu_2 =  - \frac{1}{2}  (k_B T)^{-1} \hat \gamma (u) e^{-(E_S + K)/k_B T}.
\eeq

To complete the derivation of (\ref{combi}), the statistics of the noise term
has to be specified. This requires no further information, since detailed
balance forces the stationary distribution of the continuum field $h(x,t)$ to
be governed by the Hamiltonian
\beq
\label{Hcont}
{\cal H}_c =  (\hat \gamma(u)/2) \int dx \; (\nabla h)^2,
\eeq
as can be seen from a central limit argument applied to the sum of
independent local slope variables (\ref{SOS}). A straightforward way
to ensure the stationarity of $\exp[- {\cal H}_c/k_B T]$ is to modify
the distribution functional of the white noise into the following form:
\beq
P[\eta] = \frac{e^{S[\eta]}}{Z}\,\, ,
\eeq
where $S = - (\hat \gamma(u)/2 k_B T)\int dk\int dt\,(\nu_1 + \nu_2 k^2)^{-1}
|{\hat\eta}(k,t)|^2$,
and $Z = \int {\cal D}\eta\,e^{S[\eta]}$, where ${\hat\eta}
(k,t)$ is the Fourier transform of $\eta(x,t)$.
This leads to the noise covariance
\beq
\label{noise}
\langle \eta(x,t) \eta(x',t') \rangle  = 2 (k_B T /\hat \gamma(u)) (\nu_1
+ \nu_2 \nabla^2)
\delta(x - x') \delta(t - t').
\eeq

Our derivation provides a microscopic basis for the classical theory of
Mullins \cite{mullins}, who showed that second and fourth order derivatives 
of $h$ arise from evaporation--condensation dynamics and surface diffusion
dynamics, respectively. We may further conclude that the length scale beyond
which the lower order derivative associated with evaporation--condensation
dynamics dominates is given by
\beq
\label{ellstar}
\ell^\ast \sim  \sqrt{-\nu_2/\nu_1} \sim  \exp[(1/2)(E'_S - E_S)/k_B T]
\eeq
in units of the lattice constant. Since typically the activation energies for
evaporation much exceed those for surface diffusion, $\ell^\ast$ can be quite
large at moderate temperatures. It should be noted, however, that due to
detailed balance the length scale (\ref{ellstar}) cannot appear in any
stationary, equal time correlation functions, since these only depend on the
Hamiltonian (\ref{Hcont}). Nevertheless, $\ell^\ast$ and a related time scale
will appear in the time--dependent quantities, to be discussed in subsequent
sections of this paper.
\subsection{Step--Flow Growth on Vicinal Surfaces}
\label{Schwoebel}

Technologically important deposition techniques such as molecular beam 
epitaxy
(MBE) are typically carried out at temperatures where desorption is 
negligible,
so that effectively $\nu_1 = 0$ in (\ref{nu1nu2}). However, as was first
pointed out by Villain \cite{Vil91}, under growth conditions other mechanisms
related to growth--induced surface currents \cite{Kru93} exist which
generically give rise to a second order derivative in the continuum equation.
A remarkable feature of such currents is that they can be destabilizing,
leading to $\nu_1 < 0$ in (\ref{combi}). In the present work we focus on the
kinetic roughening of a {\em stable} surface with $\nu_1 \geq 0$, and
therefore
we describe here only the simplest microscopic mechanism for the generation 
of
a positive $\nu_1$ term through the ``Schwoebel'' effect involving step edge
barriers \cite{schwoebel} (for some other mechanisms see \cite{Zangwill2}).
As in the preceding section we restrict ourselves to a one dimensional 
surface.
In two dimensions the mechanism described here gives rise to an anisotropic
Laplacian $\nu_\parallel \partial_\parallel^2 + \nu_\perp \partial_\perp^2$
in (\ref{combi}), with different coefficients $\nu_\parallel$ and $\nu_\perp$
parallel and perpendicular to the surface steps, at least one of which is
negative \cite{schimschak}.

We consider a {\em vicinal} surface with uniform step spacing $\ell$,
which is assumed to be much smaller than the diffusion length
$\ell_D$ governing the
island spacing on a {\em singular} surface \cite{Vil92}; this ensures that
island nucleation on the terraces can be neglected, and the surface grows
in the step flow mode \cite{schimschak}. Moreover we assume strong step
edge barriers which effectively suppress any interlayer transport. Under
such conditions every atom that is deposited  on a terrace attaches to
the ascending step edge, and the surface current is simply $J = F \ell/2$
(as before, $F$ denotes the deposition rate)
\cite{Vil91,Kru93,schimschak}. The coefficient $\nu_1$ is then given by the
negative derivative of $J$ with respect to the surface inclination $1/\ell$
\cite{Kru93}. This yields
\beq
\label{nu1Schwoebel}
\nu_1 = F \ell^2 /2.
\eeq
Provided the capillarity--driven surface diffusion is not too strongly 
affected
by the deposition, the expression (\ref{nu1nu2}) for $\nu_2$ is still expected
to be valid. Thus the crossover length scale (\ref{lstar}) can be estimated
as
\beq
\label{ellstar2}
\ell^\ast \sim  \ell^{-1} \sqrt{-\nu_2/F}  \sim \ell_{\rm cap}^2/\ell
\eeq
where $\ell_{\rm cap} \sim (-\nu_2/F)^{1/4}$ is a length scale gauging the
relative importance of capillarity and deposition \cite{schimschak}. For an
order of magnitude estimate, we note that $\nu_2$ can be directly measured
from the decay time of periodically modulated surface profiles. For
semiconductor surfaces a typical value is \cite{liau} $-\nu_2 \approx 1
(\mu{\rm m})^4$ per hour, implying that $\ell_{\rm cap} \approx 1000$ \AA at
a deposition rate of 1 ${\rm s}^{-1}$. This much exceeds the step spacing on
typical vicinal surfaces, and thus, as in the case of equilibrium dynamics
(Sec. \ref{Equilibrium}),
there are good reasons to expect $\ell^\ast$ to be large compared to the
lattice constant.

Of course, the most prominent effect of deposition is to provide an additional
source of ``shot noise'' fluctuations. Assuming, again, that the (volume
conserving) fluctuations due to surface diffusion are not much changed by the
deposition flux, we obtain the noise covariance
\beq
\label{noise2}
\langle \eta(x,t) \eta(x',t') \rangle  =  [F + 2 (k_B T \nu_2 /  \hat \gamma)
\nabla^2] \delta(x - x') \delta(t - t').
\eeq
A comparison between the strengths of the two components of the noise, i.e.
$F$ and $k_B T \nu_2 / \hat \gamma$, defines a
further crossover length scale
\beq
\label{noisecross}
\ell^{\ast \ast} \sim (D_{\rm coll}/F)^{1/2}
\eeq
where $D_{\rm coll} = k_0 \exp[- (E_S + K)/k_B T]$ defines, within the present
model, the {\em collective} surface diffusion coefficient \cite{Kru95}.
In contrast to the detailed balance situation of Sec. \ref{Equilibrium},
here the two crossover lengths $\ell^\ast$ and $\ell^{\ast \ast}$ need not
be equal. Their ratio is of the order
\beq
\label{ellratio}
\ell^\ast / \ell^{\ast \ast} \sim (k_B T)^{-1} \ell^{-1} \hat \gamma^{1/2}
\sim \ell^{-1} \exp(K/4 k_B T)
\eeq
where in the last step we have used the expression (\ref{stiffness}) for
small $T$.
Thus, at low temperatures $\ell^\ast \gg \ell^{\ast \ast}$. This provides some
justification for neglecting the conserved noise component in (\ref{noise2}),
as will be done throughout Sec. \ref{solutions for various quantities}.

\section{Solutions for Various Physical Quantities}
\label{solutions for various quantities}
In this section we summarize our results for the physically
interesting measures of the surface roughness of the
CLG model, for arbitrary surface dimensionalities $d \leq z$, where
$z$ is the dynamic exponent.
The physically interesting quantities that we
calculate are the surface width $w(L,t)$
and two correlation functions $G(\vec{x},t)$ and $C_s(\vec{x},t)$.
The surface width is the
size of typical height fluctuations around the mean $\bar{h} \equiv
\langle\langle (h(\vec{x},t)\rangle_x\rangle_\eta$:
\beq
w^2(L,t) \equiv \langle\langle (h(\vec{x},t) - \bar{h})^2
\rangle_x\rangle_\eta\,\, ,
\label{clgasuws}
\eeq
where $\langle\cdot\rangle_{x}$ and $\langle\cdot\rangle_{\eta}$ denote
averaging over space and noise, respectively,
and $L$ is the lateral extent of the surface, assuming periodic boundary
conditions.
The two correlation functions can be derived from the general two
point correlation function $C_g(\vec{x},t,t')$, which is defined as:
\beq
C_g(\vec{x},t,t') \equiv  \langle\langle (h(\vec{x} + \vec{x}',t +
t') - h(\vec{x}',t'))^2 \rangle_{x'}\rangle_\eta = C_g(x,t,t')\,\, .
\label{gencor}
\eeq
All the correlation functions appearing in this section are thus dependent
only on the magnitude $x = |\vec{x}|$.
The equal time correlation function $G(x,t') \equiv
C_g(x,t = 0,t')$
and the saturated correlation function $C_s(x,t)
\equiv C_g(x,t,t' \gg t_s)$, where $t_s$ is the saturation time
to be defined later. We point out that the
translational invariance of Eq. (\ref{combi})
makes the averaging over noise and space interchangeable when
the noise correlations are also translationally invariant. In this section
we assume Gaussian white noise of the form
\begin{eqnarray}
\langle \eta(\vec{x},t)\eta(\vec{x'},t') \rangle_\eta & = &
2D\delta^d(\vec{x} - \vec{x'})\delta(t - t') \\
\langle \eta(\vec{x},t) \rangle_\eta & = & 0\,\, .
\end{eqnarray}
Later on in Sec. 4 we discuss the influence
of the conserved noise component that appears in Eqs.~(\ref{noise}) and
(\ref{noise2}). Here
we also assume a flat initial condition,
$h(\vec{x},t=0) = 0$. Growth on an initially rough
surface is dealt with in Sec. \ref{applications}.

The results for the CLG equation trivially reduce to the limits
of the EW or MWV equations when $\nu_2 = 0$ or $\nu_1 = 0$,
respectively. Also, simple power counting shows that for an
{\it infinite} system, the EW behavior will always eventually dominate.
In this work, however, we are interested in investigating the crossover
time $t_c$ from MWV to EW growth. The crossover always occurs for an
{\it infinitely} large system where the
MWV behavior is dominant for early times ($t \ll
t_c$). Moreover, in a {\it finite} system for a suitable choice of the
crossover length scale $\ell^\ast$, the MWV growth can be made dominant for
all times.
As was noted already, we focus on the stable case $\nu_1 > 0$, $\nu_2 < 0$.
If $\nu_1 < 0$, the early time growth can be of MWV type but the long time
behavior of the surface is unstable.

The main results for the CLG growth equation are given in the subsequent
paragraphs. As usual \cite{reviews}, the asymptotics of the surface
correlations involve the
roughness exponent $\chi$, the dynamic exponent $z$ and the exponent ratio
$\beta = \chi/z$ which describes how the surface width increases with time.
The two limiting cases of (\ref{combi}) are characterized by the exponents
\beq
z_1 = 2, \;\;\;  \beta_1 = (2-d)/4, \;\;\; \chi_1 = (2-d)/2 \;\;\;\; 
{\rm (EW)}
\label{EWexponents}
\eeq
as was first derived in \cite{Edw82,Fam86}, and
\beq
z_2 = 4, \;\;\; \beta_2 = (4-d)/8, \;\;\; \chi_2 = (4-d)/2 \;\;\;\; {\rm
(MWV)},
\label{MWVexponents}
\eeq
compare to \cite{Wol90,DT,Golub,Lam91}. It is understood that
$\beta = \chi = 0$
implies logarithmic roughening.

The crossover time scale is given by
$t_c = |\nu_2|/\nu_1^2$ and the saturation time (for finite system size $L$)
by $t_{\mbox\scriptsize s} = L^{z_2}/(L^{z_1}\nu_1 + |\nu_2|)$.
The crossover length scale $\ell^\ast = \sqrt{|\nu_2|/\nu_1}$ was defined in
Sec. 1. In the following we use the
dimensionless scaling variables: $p_1 \equiv \nu_1 t/L^2$, $p_2 \equiv |\nu_2|
 t/L^4$, $y_1 \equiv \nu_1 t/x^2$ and $y_2 \equiv |\nu_2| t/x^4$. The dynamic
correlation lengths for the EW and MWV cases are defined as
$\xi_1 \equiv (2\nu_1 t)^{1/2}$ and $\xi_2 \equiv (2|\nu_2| t)^{1/4}$.
\noindent
\subsection{Surface Width}
%%%%%%%% SURFACE WIDTH %%%%%%%%%%%%% CLG CLG CLG CLG
%
The scaling function for the {\em surface width} is obtained
from Eq. (\ref{clgasuws}):
\beq
w^2(L,t) = \frac{2D}{\nu_1} L^{2\chi_1} F_{w}^1(p_1,p_2)
= \frac{2D}{|\nu_2|} L^{2\chi_2} F_{w}^2(p_1,p_2)\,\, ,
\eeq
where
\beq
F_{w}^j(p_1,p_2) = \frac{\Omega_{d-1}}{2(2\pi)^d} \int_1^\infty
dk\,k^{d-1}[1 - e^{-2(p_1k^2 + p_2k^4)}]
\frac{p_j}{p_1k^2 + p_2k^4}\,\, ,
\label{suwe}\nonumber
\eeq
where $\Omega_{d-1}$ is the surface area of a $d$ dimensional sphere,
and $j = 1,2$. We obtain the following power law behavior in the different
time regimes:
\beq
\begin{array}{ll}
\mbox{If $p_1 \gg p_2$, then} & \\
   & w^2 \propto \left\{\begin{array}{ll}
     L^{2\chi_1}\,\, ,  &\mbox{for $t \gg t_{s}\,\, ;$} \\
     t^{2\beta_1}\,\, , &\mbox{for $t_c \ll t \ll t_{s}\,\, .$} \\
     t^{2\beta_2}\,\, , &\mbox{for $t \ll t_c\,\, .$}
                        \end{array}\right. \\
\mbox{If $p_1 \ll p_2$ then} & \\
   & w^2 \propto \left\{\begin{array}{ll}
     L^{2\chi_2}\,\, ,  &\mbox{for $t \gg t_{s}\,\, ;$} \\
     t^{2\beta_2}\,\, , &\mbox{for $t \ll t_{s}\,\, .$}
                        \end{array}\right.
\end{array}
\eeq
The main result here is that
for a {\it finite} system the MWV behavior can dominate for all times,
including the saturated regime. This is shown in Fig. $1$,
where it can be seen that even for $p_1=p_2$,
the EW region can be made to vanish by choosing $t_s= t_c$.
\subsection{Equal Time Correlation Function}

%%%%%%%%%%%%% EQUAL TIME CORRELATION FUNCTION %%%%%%%%%%% CLG CLG CLG
%
%
\noindent
The scaling function for the {\em equal time correlation
function} is obtained from Eq. (\ref{gencor}). Hence,
\beq
G(x,t) = \frac{2D}{\nu_1}x^{2\chi_1} F_G^1(y_1,y_2)
= \frac{2D}{|\nu_2|}x^{2\chi_2} F_G^2(y_1,y_2)\,\, ,
\label{equu}
\eeq
where
\beq
F_G^j(y_1,y_2) = \frac{1}{(2\pi)^d}\int_\Omega\int_0^\infty dk\,k^{d-1}
[1 - e^{-2(y_1k^2 + y_2k^4)}][1 - \cos(k\alpha)]
\frac{y_j}{y_1k^2 + y_2k^4}\,\, ,
\label{gewe}\nonumber
\eeq
and $j = 1,2$. The notation $\int_\Omega$ means angular integration, and
$\alpha \equiv \cos({\vec k}, {\vec x})$. We obtain the following power law
behavior in the different regimes:
\beq
\begin{array}{ll}
\mbox{If $p_1 \gg p_2$ then} & \\
   & G(x,t) \propto \left\{\begin{array}{ll}
     L^{2\chi_1}\, ,  &\mbox{for $x = {\cal O}(L)$ and} \\
                      &\mbox{$t \gg t_{s}$;} \\
     x^{2\chi_1}\, ,  &\mbox{for $\ell^\ast \ll x \ll L$ and} \\
                      &\mbox{$t \gg t_{s}$;} \\
     x^{2\chi_2} (L/x)^{2\chi_1}\, ,  &\mbox{for $x \ll
                      \ell^\ast \ll L$ and} \\
                      &\mbox{$t \gg t_{s}$;} \\
     x^{2\chi_1}\, ,  &\mbox{for $\xi_1 \gg x \gg \ell^\ast$
and} \\
                      &  \mbox{$t_c \ll t \ll t_{s}$;} \\
     t^{2\beta_1}\, , &\mbox{for $\ell^\ast \ll \xi_1 \ll x$
and} \\
                      &  \mbox{$t_c \ll t \ll t_{s}$;} \\
     x^{2\chi_2} (\xi_2/x)^{2\chi_1}\, ,  &\mbox{for $x \ll
                    \xi_2 \ll \ell^\ast$ and} \\
                      &\mbox{$t \ll t_c$;} \\
     t^{2\beta_2}\, , &\mbox{for $x \gg \xi_2$ and $t \ll t_c$.}
                        \end{array}\right. \\
%%%%%%%
\mbox{If $p_1 \ll p_2$ then} & \\
   & G(x,t) \propto \left\{\begin{array}{ll}
     L^{2\chi_2}\, ,  &\mbox{for $x = {\cal O}(L)$ and} \\
                      &\mbox{$t \gg t_{s}$;} \\
     x^{2\chi_2} (L/x)^{2\chi_1}\, ,  &\mbox{for $x \ll L$ and
                      $t \gg t_{s}$;} \\
     x^{2\chi_2} (\xi_2/x)^{2\chi_1}\, ,  &\mbox{for $x\ll
                      \xi_2$ and $t \ll t_{s}$;} \\
     t^{2\beta_2}\, , &\mbox{for $x \gg \xi_2$ and $t \ll t_{s}$.}
                        \end{array}\right.
\end{array}
\eeq
To summarize, the correlation function can exhibit both EW and MWV
scaling behavior for $p_1 \gg p_2$, while in the opposite case
we find ``anomalous'' scaling behavior
\beq
G \sim x^{2\chi_2} (L/x)^{2\chi_1}, \;\;\; G \sim x^{2\chi_2}
(\xi_2/x)^{2 \chi_1}
\label{anomal}
\eeq
of the kind characteristic of the MWV equation \cite{Ama93}
as well as certain nonlinear models \cite{Sar94,schroeder,Kru94}.
The scaling (\ref{anomal}) is anomalous in the sense that $G$
at fixed $x$ has no finite limit for $L \to \infty$ and $\xi_2 \to \infty$;
this implies the appearance of arbitrarily large height gradients and
is associated with the fact that the MWV roughness exponent $\chi_2 > 1$
for $d < 2$, compare to (\ref{MWVexponents}). Note that the increase of
$G$ with $x$ in (\ref{anomal}) is governed neither by $\chi_1$ nor by
$\chi_2$, but by an anomalous roughness exponent $\tilde \chi =
\chi_2 - \chi_1 = 1$.

\subsection{Saturated Correlation Function}
%%%%%%%%%%%%% SATURATED CORRELATION FUNCTION %%%%%%%%%% CLG CLG CLG
%
\noindent
The scaling function for the {\em saturated correlation
function} is obtained from Eq. (\ref{gencor}). Hence,
\beq
C_s(x,t) = \frac{2D}{\nu_1}x^{2\chi_1} F_{C}^1(y_1,y_2)
= \frac{2D}{|\nu_2|}x^{2\chi_2} F_{C}^2
(y_1,y_2)\,\, ,
\eeq
where
\beq
F_{C}^j(y_1,y_2) = \frac{1}{(2\pi)^d}\int_\Omega\int_0^\infty dk\,k^{
d-1}[1 - e^{-(y_1k^2 + y_2k^4)}\cos(k\alpha)]
\frac{y_j}{y_1k^2 + y_2k^4}\,\, , \nonumber
\eeq
and $j = 1,2$. For simplicity, we only give the results
in the limits where either $t=0$ or $x=0$. In the first case,
the power law behavior of $C_s(x,0)$ is given by:
\beq
\begin{array}{ll}
\mbox{If $p_1 \gg p_2$ then} & \\
   & C_s(x,0) \propto \left\{\begin{array}{ll}
     L^{2\chi_1}\, ,  &\mbox{for $x = {\cal O}(L)$;} \\
     x^{2\chi_1}\, ,  &\mbox{for $\ell^\ast \ll x \ll L$;} \\
     x^{2\chi_2} (L/x)^{2\chi_1}\, ,  &\mbox{for $x \ll
                     \ell^\ast \ll L$.}
                         \end{array}\right. \\
\mbox{If $p_1 \ll p_2$ then} & \\
   & C_s(x,0) \propto \left\{\begin{array}{ll}
     L^{2\chi_2}\, ,  &\mbox{for $x = {\cal O}(L)$;} \\
     x^{2\chi_2} (L/x)^{2\chi_1}\, ,  &\mbox{for $x \ll L$.}
                         \end{array}\right.
\end{array}
\eeq
The behavior of $C_s(0,t)$ is given by:
\beq
\begin{array}{ll}
\mbox{If $p_1 \gg p_2$ then} & \\
   & C_s(0,t) \propto \left\{\begin{array}{ll}
     L^{2\chi_1}\, ,  &\mbox{for $t \gg t_{s}$;} \\
     t^{2\beta_1}\, , &\mbox{for $t_c \ll t \ll t_{s}$;} \\
     t^{2\beta_2}\, , &\mbox{for $t \ll t_c$.}
                        \end{array}\right. \\
\mbox{If $p_1 \ll p_2$ then} & \\
   & C_s(0,t) \propto \left\{\begin{array}{ll}
     L^{2\chi_2}\, ,  &\mbox{for $t \gg t_{s}$;} \\
     t^{2\beta_2}\, , &\mbox{for $t \ll t_{s}$.}
                        \end{array}\right.
\end{array}
\eeq
As in Sec. 3.2, both EW and MWV behavior are found in the different
regimes.
\section{Applications}
\label{applications}
The goal of the present section is to illustrate how linear growth equations
can be used to address experimentally relevant questions about kinetic
roughening that have so far received little or no theoretical attention.
Specifically, we discuss the contribution of the substrate roughness to the
width of a growing surface, and the detailed form
of the spatial height correlation function.
\subsection{Effects of Substrate Roughness}
\label{Substrate}
In the real world, thin films are rarely deposited onto a perfect, atomically
flat substrate. Consequently every experimental investigation of kinetic
roughening has to deal with the substrate contribution to the roughness of the
film surface. To the extent that this problem has been addressed at all, it is
usually assumed \cite{krim}
that the substrate gives an additive, constant contribution
to the variance of the height fluctuations (the square of the width), as
\begin{equation}
\label{widthinitial}
w^2(t) = w_i^2 + w_G^2(t)
\end{equation}
where $w_i$ is the width of the substrate surface and $w_G$ denotes the growth
induced contribution (throughout this section the lateral system size is taken
to be infinite and therefore the dependence on $L$ is suppressed). This simple
ansatz ignores the fact that the memory of the initial roughness is
lost during the growth process, as the growing film successively covers up
the features of the substrate. The short wavelength features
are preferentially suppressed (compare to a layer of snow covering a rugged
landscape), an effect of much importance for the scattering from multilayer
films \cite{salditt}.

In the following we show that (i) within the framework of linear
growth equations, the superposition ansatz (\ref{widthinitial}) is justified,
however (ii) the substrate contribution $w_i$ becomes time dependent and
decreases with $t$ in a manner governed by the ratio of the substrate
correlation length $\xi_s$ to the correlation length $\xi(t) \sim t^{1/z}$ of
the growth process; for long times, $\xi \gg \xi_s$, we find
\beq
\label{wi0}
w_i(t) \sim w_i(0) (\xi_s/\xi)^{d/2}.
\eeq
The fact that the total width is the sum of a decreasing and an
increasing part entails the somewhat counterintuitive possibility that $w(t)$
may initially {\em decrease} with increasing film thickness, as has been
observed in recent experiments \cite{uwe,frank}.

To justify statements (i) and (ii), we assume that
the surface has been grown from time $t = -t_0$ to $t=0$ driven by
some initial noise $\eta_-$ in such a way that the height--height
correlation function is given by
\beq
\langle \hat{H}_0(\vec{k})\hat{H}_0(\vec{k'}) \rangle_{\hat{\eta}_-} =
f(\vec{k},\vec{k'})\,\, ,
\eeq
where $\hat{H}_0(\vec{k}) \equiv \hat{H}_0(\vec{k},t=0)$, and
the average has been taken over all configurations
created by the noise $\eta_-$. We note that for $t>0$, $\eta_-\equiv 0$.
At $t=0$, a new growth process with noise $\eta$ is turned on. Let us
denote the full solution of the CLG equation by $\hat{H}(\vec{k},t)$
with the initial condition $\hat{H}_0(\vec{k})$.
Then,
\beq
\hat{H}(\vec{k},t) = \hat{H}_0(\vec{k}) e^{-a(\vec{k})t} + \hat{h}(
\vec{k},t)\,\, ,
\eeq
where $a(\vec{k}) \equiv \nu_1 k^2 + |\nu_2| k^4$, and $\hat{h}(\vec{k},t)$
denotes the solution of the CLG equation with a flat initial condition.
Calculating the correlation function with respect to both the old
noise $\eta_-$ (from $t=-t_0$ to $t=0$) and a new noise $\eta$ (for
$t>0$) gives
\begin{eqnarray}
\lefteqn{ \langle
\langle \hat{H}(\vec{k},t)\hat{H}(\vec{k'},t) \rangle_{\hat{
\eta}}\rangle_{\hat{\eta}_-} = } \nonumber \\
& & f(\vec{k},\vec{k'})e^{-[a(\vec{k}) + a(\vec{k'})]t} + \langle
\hat{h}(\vec{k'},t)\hat{h}(\vec{k},t) \rangle_{\hat{\eta}}\,\, ,
\label{furu}
\end{eqnarray}
given that either $\langle \hat{H}_0(\vec{k}) \rangle_{\hat{\eta}_-} = 0$,
or $\langle \hat{h} (\vec{k},t) \rangle_{\hat{\eta}} = 0$,
which makes the crossterms disappear. The influence of the initial conditions
vanishes exponentially fast in the Fourier space but not necessarily in the
real space. To see this we consider the following example. At time $t=-t_0$ 
the
surface is flat. Then we switch on the beam and let the surface evolve until
$t=0$ driven by the CLG growth dynamics (the noise $\eta_-$ is white). For $f$
we get
\begin{eqnarray}
f(\vec{k},\vec{k'}) & = & \langle \int_{-t_0}^0 d\tau\,\hat{\eta}_-(\vec{k},
\tau) e^{a(\vec k)\tau} \int_{-t_0}^0 d\tau'\,\hat{\eta}_-(\vec{k'},\tau')
e^{a(\vec k)\tau'}
\rangle_{\hat{\eta}_-} \nonumber \\
                    & = & \frac{2D}{(2\pi)^d}\frac{1 - e^{-2a(\vec k)t_0}}
{2a({\vec k})}
\delta^d(\vec{k} + \vec{k}')\,\, .
\end{eqnarray}
Integrating the first term on the right hand side of (\ref{furu}) with
respect to $\vec{k}$ shows that the initial roughness decays slowly, as
$w_i^2 \sim t_0/t$, in the limit $t \to \infty$, when the substrate
dimension $d=2$. This corresponds precisely to (\ref{wi0}) with EW scaling,
$\xi_s \sim t_0^{1/2}$ and $\xi \sim t^{1/2}$.

In the preceding we dealt with the full CLG equation.
Next we will focus on calculating the time dependent substrate surface width
$w_i^2(t)$ for the special cases $\nu_2 = 0$ (EW equation, $z=2$) and $\nu_1 =
0$ (MWV equation, $z=4$) of (\ref{combi}). Moreover, we assume that the
substrate surface has been grown by either EW or MWV dynamics, so that its
correlations can be described by the Fourier amplitudes
\begin{equation}
\label{Fourierinitial}
\langle \vert \hat H_0(\vec k) \vert^2 \rangle_{\hat{\eta}_-} =
\frac{A_0}{k^{z_s}} [1 - e^{-(k \xi_s)^{z_s}}]
\end{equation}
where $\xi_s$ is the substrate correlation length,
and $z_s = 2$ or $z_s = 4$ for surfaces generated by EW or MWV dynamics,
respectively.
The roughness exponent of the substrate is $\chi_s = (z_s-d)/2$.
Solving (\ref{combi}) subject to the initial condition
$\hat H_0(\vec k)$ one obtains
\begin{equation}
\label{decomp}
\langle \langle \vert \hat H(\vec k,t) \vert^2 \rangle_{\hat{\eta}}
\rangle_{\hat{\eta}_-} = \langle \vert \hat H_0(\vec k) \vert^2
\rangle_{\hat{\eta}_-} e^{-(k \xi(t))^z} + \frac{D}{\nu}
\frac{1}{k^z}[1 - e^{-(k \xi(t))^z}]
\end{equation}
where $z$ is the dynamic exponent of the growth process and $\xi(t) = (2 \nu
t)^{1/z}$ its correlation length. Since the variance $w^2$ is
obtained by integrating (\ref{decomp}) over $k$, the decomposition
(\ref{widthinitial}) is valid and the substrate contribution at time $t$ is
given by
\begin{equation}
\label{wi1}
w_i^2(t) = \Omega_{d-1}
A_0 \int_0^{k_{\rm max}} dk \; k^{d-1-z_s} [1 - e^{-( k \xi_s)^{z_s}}] \;
e^{-(k \xi)^z}\,\, .
\end{equation}
The upper cutoff $k_{\rm max}$ is of the order of the inverse lattice
constant; in the following we assume that $\xi_s$ and $\xi$ are large compared
to $1/k_{\rm max}$ and set $k_{\rm max} = \infty$ in (\ref{wi1}).

Substituting $q = (k \xi_s)^{z_s}$ in (\ref{wi1}) yields
\begin{equation}
\label{wi2}
w_i^2(t) = (\Omega_{d-1} A_0 /z_s) \xi_s^{z_s - d}
\int_0^\infty dq \; q^{d/z_s -2} ( 1 - e^{-q})
e^{-\tau q^{z/z_s}}
\end{equation}
where the dimensionless time variable $\tau = (\xi/\xi_s)^z$ has been
introduced (recall that $\xi \sim t^{1/z}$). The integral can be explicitly
computed if $z = z_s$; details are given in Appendix A.
To analyze (\ref{wi2}) in the general case, let us first assume that
$z_s > d$, i.e. $\chi_s > 0$, which covers MWV substrates in $d=1$ and
$d=2$, and EW substrates in $d=1$.

This ensures that
the integral over $q^{d/z_s -2}$ converges at infinity, and hence the factor
$\exp(-\tau q^{z/z_s})$ can be dropped when $\tau \ll 1$. The width is
therefore independent of $\tau $ for $\tau \ll 1$. Physically this simply
reflects the fact that, for a surface with power law roughness, $\chi_s > 0$,
the width is dominated by the
long wavelength fluctuations with wavelengths of the order of $\xi_s$. At time
$t$ initial fluctuations of wavelengths up to the correlation length $\xi(t)$
have been eliminated, thus the substrate contribution to the width decreases
appreciably only when $\xi \approx \xi_s$ or $\tau \approx 1$.

For $z_s = d$ (i.e. $\chi_0 = 0$, for example an EW substrate in $d=2$)
the integral over $q^{d/z_s -2}$ diverges logarithmically at large $q$.
The factor $\exp(-\tau q^{z/z_s})$ then has to be retained, and one finds
that $w_i(\tau)^2 \sim \ln (1/\tau)$ for $\tau \ll 1$; of course this
behavior is valid only for $\tau > (k_{\max} \xi_s)^{-z}$, since the
initial roughness $w_i(0)^2 \approx \Omega_{d-1} A_0 \ln (k_{\max} \xi_s)$.
Finally, in the
(somewhat academic) case $z_s < d$, $w_i(\tau)$ decreases as a power law also
for $\tau \ll 1$, as $w_i(\tau)^2 \sim \xi^{-(d-z_s)}$, see Appendix A.

The behavior for large $\tau$ can be discussed independent of $z_s$. For $\tau
\gg 1$ the integral (\ref{wi2}) is effectively cut off at $q \approx \tau^{
-z_s/z} \ll 1$, therefore we can set $1 - e^{-q} \approx q$ in the integrand,
and it follows by rescaling that
\begin{equation}
\label{wigen2}
w_i^2(\tau) \approx A_0 \xi_s^{z_s - d} (\xi_s/\xi)^d \sim t^{-d/z}.
\end{equation}
This result has a simple interpretation. For $\xi \gg \xi_s$ the growth
process averages over a large number $N = (\xi/\xi_s)^d$ of domains in which
the initial fluctuations are statistically independent. The height variance of
each domain is of the order $A_0 \xi_s^{z_s - d}$, and averaging over $N$
domains reduces the variance by a factor of $1/N$. If the substrate has power
law roughness ($\chi_s > 0$), then $w_i(0)^2 \approx A_0 \xi_s^{z_s-d}$ and
(\ref{wigen2}) is identical to (\ref{wi0}).

The last argument is valid also for completely general initial
conditions characterized by a correlation function of the form
\begin{equation}
\label{Fourierinitial2}
\langle \vert \hat H_0(\vec k) \vert^2 \rangle_{\hat{\eta}_-} =
\frac{\Omega_{d-1} A_0}{k^{d+2\chi_0}} g(k \xi_s)
\end{equation}
where $\chi_0$ denotes the roughness exponent of the substrate and the scaling
function $g(s)$ satisfies $g(s \to \infty) = 1$, $g(s \to 0) = 0$. The
expression (\ref{wi2}) generalizes to
\begin{equation}
\label{wi3}
w_i^2(\tau) = \Omega_{d-1} A_0 \xi_s^{2 \chi_0} \int_0^\infty dq \; q^{-(1 +
2\chi_0)} g(q) e^{-\tau q^z}.
\end{equation}
To extract the behavior for large $\tau$ we need to know how $g(s)$ vanishes
for small $s$. This is fixed by requiring that (\ref{Fourierinitial2}) should
have a finite limit for $k \to 0$ (this limit gives rise to the center of mass
fluctuations of the surface, see \cite{1/f}). Consequently $g(s) \sim s^{d +
2\chi_0}$, and (\ref{wi3}) decays as
\begin{equation}
\label{wigen3}
w_i^2(\tau) \approx \Omega_{d-1} A_0 \xi_s^{2 \chi_0} \tau^{-d/z} \approx
w_i^2(0) (\xi_s/\xi)^d
\end{equation}
in accordance with the heuristic argument.

In summary, we have shown that, under rather general conditions, the substrate
contribution $w_i$ to the width of a rough growing surface remains essentially
constant as long as the correlation length $\xi$ of the growth process is
smaller than the substrate correlation length $\xi_s$, and that it decreases
according to (\ref{wi0}) for $\xi \gg \xi_s$. While the functional form of
the transition between the two regimes is not analytically accessible in
general, a useful interpolation formula is
\beq
\label{interpolation}
w_i(t)^2 = w_i(0)^2 (1 + t/t_s)^{-d/z}
\eeq
with a fit parameter $t_s$,
which has already been employed in the analysis of experimental data
\cite{frank}. In Fig. 2 the decay of the substrate width is illustrated
for two special cases, and the formula (\ref{interpolation}) is compared to
the exact expression derived in Appendix A.

\subsection{Shape of the Height Correlation Function}
\label{Correlation}
The dynamic scaling hypothesis of kinetic roughening theory \cite{reviews}
states that the
height--height correlation function should have the scaling form
\begin{equation}
\label{Cscal}
C(x,t) \equiv \langle h(\vec x' + \vec x,t) h(\vec x',t) \rangle = w^2(t)
F(x/\xi(t))
\end{equation}
with $w^2 = C(0,t)$, hence $F(0) = 1$, and $\xi \sim t^{1/z}$. To date almost
all theoretical work has focused on the behavior of the height difference
correlation function
\begin{eqnarray}
\label{Gscal}
G(x,t) & = & \langle (h(\vec x' + \vec x,t) - h(\vec x',t))^2 \rangle
\nonumber \\
       & = & 2 [C(0,t) - C(x,t)] = 2 w^2[1 - F(x/\xi(t))]
\end{eqnarray}
for $x \ll \xi$, i.e. the short distance behavior
\begin{equation}
\label{Fshort}
F(s) \approx 1 - {\cal O}(s^{2 \chi})\,\, ,\;\;\; s \to 0\,\, ,
\end{equation}
of the scaling function. In
contrast, the overall shape of $F$ or, in particular, the way it decays for
large arguments has not been addressed theoretically,
although a considerable amount of
empirical information is available \cite{tong94}.
This question is of some
experimental importance, since the interpretation of scattering data from
rough surfaces typically requires a model for the entire correlation function
\cite{Pal93}. A widely used form for $F$ is \cite{Sin88}
\begin{equation}
\label{Fmodel}
F(s) = \exp(-s^{2\chi})\,\, ,
\end{equation}
which assumes a simple relation between large distance and short distance
behaviors. Here we show by explicit calculations within the linear theory
that the large distance decay of $F$ involves a new exponent not simply
related to $\chi$, and that moreover the decay can be oscillatory rather
than monotonic.

First we make some general remarks concerning the correlation function for the
CLG model. The entire correlation function valid for all ranges can be
rewritten for $d=2$ as (by substitution $k' = k/\xi$ in Eq.~(\ref{equu}))
\beq
G(\frac{x}{\xi},\frac{\xi_1}{\xi},\frac{\xi_2}{\xi}) = \frac{D}{\pi\nu_1}
\frac{\xi_1^2}{\xi^2} \int_0^\infty dk\,k [1 - J_0(k x/\xi)]\frac{[1 -
e^{-(\frac{\xi_1^2}{\xi^2} k^2 + \frac{\xi_2^4}{\xi^4} k^4)}]}{\frac{
\xi_1^2}{\xi^2} k^2 + \frac{\xi_2^4}{\xi^4} k^4}\,\, ,
\label{geeku}
\eeq
where $\xi_1 \equiv (2\nu_1 t)^{1/2}$ and $\xi_2 \equiv (2|\nu_2| t)^{1/4}$.
As far as the CLG model can be considered to describe the physics correctly
either Eq.~(\ref{geeku}) or Eq.~(\ref{equu}) could be used directly as fitting
functions, with the parameters ($D$, $\nu_1$, $\nu_2$) or ($A \equiv D/
(2\pi\nu_1)$, $\xi_1$, $\xi_2$) to be fitted. The amplitude is set by
parameters $D$ or $A$. The correlation length $\xi(t)$ in Eq.~(\ref{geeku})
corresponds to the experimentally observable correlation length.
We can numerically determine the dependence of
$\xi$ on $\xi_1$ and $\xi_2$ by using the relation $C(x=\xi)/C(x=0) = e^{-1}$,
which is the usual definition of the correlation length. The surface
$\xi(\xi_1,\xi_2)$ is plotted in Fig. $3$. A least squarest fit for a
set of solutions $\xi = \xi(\xi_1,\xi_2)$ yields
\beq
\xi_p \equiv \alpha_1\xi_1 + \alpha_2\xi_2 = 0.08533\xi_1 + 1.60573\xi_2\,\, .
\label{xiku}
\eeq
In addition, numerical studies show
that $\xi$ follows the scaling relation $\xi(a\xi_1,a\xi_2) = a\xi(\xi_1,
\xi_2)$. Our choice $\xi_p \equiv \alpha_1\xi_1 + \alpha_2\xi_2$ is a simple
case which quite accurately satisfies the definition for the correlation
length.

Unlike in the case of the heuristic correlation function (\ref{Fmodel}), the
requirement $x \ll \xi$ alone is not sufficient to determine the scaling
behavior of $G$ for CLG model with a fixed exponent $\chi$. The scaling
analysis of Sec. \ref{solutions for various quantities} shows that the
scaling of the correlation function depends on two dimensionless quantites:
$\xi_1/\xi_2 \sim (t/t_c)^{1/2}$ and $\xi_1 x/\xi_2^2 \sim x/\ell^\ast$. If
$t/t_c \gg 1$, $x/\ell^\ast \gg 1$, and $x \ll \xi$, then $G \propto
x^{2\chi_1(d=2)}$ and $\chi \approx \chi_1(d=2) = 0$. If $t/t_c \ll 1$,
$x/\ell^\ast \ll 1$, and $x \ll \xi$, then $G \propto x^{2\chi_2(d=2)}
\ln(\xi_2/x)$ and thus $\chi \approx \chi_2(d=2) = 1$.

Let us next determine the behavior of the correlation function for
large values of the ratio $x/\xi \equiv s$. Rather than trying to present
the full solution for the CLG model we shall limit ourselves to the one
and two dimensional EW and MWV models. For the $d=1$ EW case,
the evaluation of $C$ is straightforward. We have
\beq
\partial_t C(x,t) = \frac{2Ds}{\pi x} \int_0^\infty dk\,e^{\imath ks -
k^2}\,\, .
\eeq
This is a Gaussian integral which can be done. The integration over $t$ is
also standard. We obtain the result:
\beq
C(x,t) = w^2(t)F(s)\,\, ,
\eeq
with
\begin{eqnarray}
w^2(t) & = & \frac{D\xi_1}{\nu_1\sqrt{\pi}}\,\, ;\\
F(s)   & = & \frac{1}{2}\int_0^1 dy\,y^{-1/2}e^{-s^2/(4y)} \to 2 s^{-2}
e^{-s^2/4}
\end{eqnarray}
as $s\to\infty$. The behavior of the scaling function $F$ is in striking
contradiction to the conventional ansatz (\ref{Fmodel}), which would lead one
to expect an exponential decay for $\chi = 1/2$.

The $d=2$ EW case is also simple to calculate:
\beq
\partial_t C(x,t) = \frac{D}{\pi}\int_0^\infty dk\,kJ_0(kx)e^{-(\xi_1
k)^2}\,\, .
\eeq
Integrating the result from $0$ to $t$ yields \cite{Nat92}
\beq
C(s) = \frac{D}{4\pi\nu_1} \int_{s^2/4}^\infty dy\,y^{-1}e^{-y}
\equiv \frac{D}{4\pi\nu_1} E_1(s^2/4)\,\, ,
\eeq
where $E_1$ is an exponential integral.
The asymptotic power law expansion of $E_1$ reveals that as a function of
$s$ the leading term is equivalent to the $d=1$ case: $E_1(x) \sim x^{-1}
e^{-x}(1 - x^{-1} + 2x^{-2} - \cdots)$, hence $F(s) \sim s^{-2} e^{-s^2/4}$.

In the $d=1$ MWV case we confront the integral \cite{Col95}
\beq
\partial_t C(x,t) \propto \xi_2^{-1}\int_{-\infty}^\infty dk\,e^{\imath sk -
k^4} \equiv \xi_2^{-1}\int_{-\infty}^\infty dk\,e^{-f(k)}\,\, .
\eeq
The behavior of the leading asymptotic behavior can be evaluated using the
method of stationary phase \cite{Dingle}. In the limit $s\to\infty$ only
the neighborhoods of saddle points contribute to the integral. They are
determined from the relation $\partial_k f = 0$:
\beq
k_0 \equiv (\frac{s}{4})^{1/3}e^{\imath \pi/6}\,\,\, ,\,\,\,k_1 \equiv (\frac{
s}{4})^{1/3}e^{\imath 5\pi/6}\,\,\, ,\,\,\,k_2 \equiv (\frac{s}{4})^{1/3}e^{
\imath 3\pi/2}\,\, .
\eeq
Removal of each saddle point to the origin results in a quadratic dependence
of $f(k)$ on $k$: $f(k) - f(k_i) \to b_i k^2$ as $k\to 0$, $i =
0,1,2$. Thus, in the leading approximation we retain only the Gaussian
part of the integrand:
\beq
\partial_t C(x,t) \propto \xi_2^{-1} b_i^{-1/2}e^{-f(k_i)}\,\, .
\eeq
Furthermore, we note that the contributions coming from distinct saddle
points are additive if the paths of integration are properly chosen in the
complex plane. Despite that
the individual contributions coming from points $k_0$
and $k_1$ are complex, their sum is real.
The point $k_2$ has to be discarded
as it would cause an exponential blow--up. Finally, we obtain
\beq
C(x,t) \sim \frac{\xi_2^{3}}{|\nu_2|}s^{-5/3}\,e^{-a_1 s^{4/3}} [a_1
\sin(a_2 s^{4/3}) + a_2\cos(a_2 s^{4/3})]\,\, ,
\eeq
where $a_1 \equiv (3/8)4^{-1/3}$ and $a_2 \equiv (3\sqrt{3}/8)4^{-1/3}$.
The correlation function $C$ is an oscillating function for large values
of the argument $s$ and can assume negative values as well.

For the $d=2$ MWV case we use the same method as in the previous case:
\beq
\label{CMWV2}
\partial_t C(x,t) = \frac{2D}{(2\pi)^2}\int_{-\infty}^\infty dk_1
\int_{-\infty}^{\infty} dk_2\,e^{-2\xi_2^4(k_1^2 + k_2^2)^2 + \imath(k_1 x_1
+ k_2 x_2)}\,\, .
\eeq
To determine the saddle points we set $\partial_{k_1} f = \partial_{k_2}
f = 0$, where $f$ denotes the expression in the exponential of (\ref{CMWV2}).
This yields
\begin{eqnarray}
\partial_{k_1} f = 0\ &;&\ \partial_{k_2} f = 0\,\, , \\
\Rightarrow\,\,\,4\xi_2^4 k_1^3 + 4\xi_2^4 k_1 k_2^2 -\imath x_1 =
0\ &;&\ 4\xi_2^4 k_2^3 + 4\xi_2^4 k_2 k_1^2 -\imath x_2 = 0\,\, ,
\end{eqnarray}
which can be solved with the ansatz $k_1 = ax_1$, $k_2 = ax_2$. Again we get
three saddle points, one of which corresponds to an exponential divergence.
Summing up the contributions of the remaining two and integrating over time
gives a real answer:
\beq
C(s) \sim \frac{\xi_2^2}{|\nu_2|} s^{-2}\,e^{-a_1 s^{4/3}}
\cos(a_2 s^{4/3})\,\, .
\eeq
As before, an oscillating function results.

Summarizing, it appears that the large distance decay of the scaling function
$F(s)$ in (\ref{Cscal}) generally has the form of a `squeezed' exponential
decorated by a subleading power law,
\beq
\label{genscal}
F(s) \approx s^{-\gamma} e^{-c s^\delta}, \;\;\; s \to \infty,
\eeq
where the amplitude $c$ can be real (as in the EW equation) or complex (as in
the MWV equation); in the latter case the decay is modulated by oscillations.
In contrast to what is suggested by the heuristic function (\ref{Fmodel}),
the decay exponent has no direct relation to the roughness exponent $\chi$;
rather, in the examples treated here, $\delta$ seems to be characteristic
of the surface relaxation process but independent of dimensionality.
In Appendix B we give an alternative derivation for $d=1$ which suggests
that $\delta = z/(z-1)$ and $\gamma = 1 + \delta/2$ for the generalized
linear equation
\beq
\label{genlinear}
\partial_t h = - (- \partial_x^2)^{z/2} h + \eta
\eeq
with integer values of $z/2$.

It is worth pointing out that a rapid, exponential--like decay of the
correlation function should be expected only for growth processes that are
{\em local} in space, such as (\ref{genlinear}), for which the Fourier
transform of $C(\vec{x},t)$ is an analytic function of $\vec{k}$ for small
$\vert \vec{k} \vert$. A simple example of a nonlocal kinetic roughening
process is diffusion limited erosion \cite{dle}, formally the generalization
of (\ref{genlinear}) to $z=1$, for which it can be shown that in $d=1$
\beq
\label{Cdle}
C(x,t) =  \frac{1}{2\pi} \frac{D}{\nu} \ln (1 + \xi^2/x^2)
\eeq
for distances larger than the lattice spacing $a$; at $x \approx a$,
$C \approx w^2 \approx (D/\nu) \ln (\xi/a)$, since $\chi = (1-d)/2 = 0$ in
$d=1$ \cite{dle}. As usual, the correlation length is $\xi = 2 \nu t$.
At large distances (\ref{Cdle}) decays as a power law, $C \sim (\xi/x)^2$,
reflecting the nonlocal nature of the interface dynamics.

\subsection{Effects of Noise}
\label{effects of noise}
We have shown in Sec. \ref{derivation of linear growth model} that the
correlation function for the noise typically consists of two components,
\beq
\label{2noise}
\langle \eta(\vec{x},t) \eta(\vec{x}',t') \rangle  = 2(D_1 - D_2
\nabla^2) \delta(\vec{x} - \vec{x}') \delta(t - t')
\eeq
where the first, white noise component arises from evaporation and
deposition, while the second, conserved component reflects the
thermal fluctuations in the surface current \cite{Vil91}. In Fourier space the
correlator (\ref{2noise}) reads
\beq
\label{Fouriernoise}
\langle \tilde \eta (\vec k,\omega) \tilde \eta (\vec{k}',\omega',)
\rangle_{ \tilde \eta} = (2\pi)^{-d}(2D_1 + 2D_2 k^2)
\delta^d(\vec{k} + \vec{k}')\delta(\omega + \omega')\,\, ,
\eeq
where $\tilde \eta$ denotes the Fourier transform with respect to both space
and time. So far we have considered the behavior of the surface on scales
large compared to the noise crossover scale $\ell^{\ast \ast} \sim
(D_2/D_1)^{1/2}$, where the conserved part can be neglected. In the opposite
regime $x \ll \ell^{\ast \ast}$ we can set $D_1 = 0$ and analyze the
behavior generated only by diffusion noise. The corresponding power laws
are easily evaluated: Because the $D_2$ correlator in (\ref{Fouriernoise})
is of the same form as for the white noise
apart from the factor $k^2$, the results presented in Sec.
\ref{solutions for various quantities} hold if we replace $d$ by $d+2$.

In Ref. \cite{Med89} it was pointed out that the effect of fast degrees of
freedom may sometimes be represented by means of a coloured noise term
$\eta_c$, i.e. the noise has power law correlations. Through renormalization
group analysis it was shown that power law correlated noise changes the
scaling exponents for the KPZ equation. Now we shall extend our analysis of
Eq.~(\ref{combi}) to include power law correlated noise as well. We assume
that the noise has long distance correlations of the form
\beq
\langle \eta_c(\vec{x},t)\eta_c(\vec{x'},t') \rangle_{\eta_c} =
|\vec{x} - \vec{x'}|^{2\rho - d}|t - t'|^{2\varrho-1}\,\, ,
\label{noisespec}
\eeq
with $0 < \rho < d/2$ and $0 < \varrho < 1/2$. Taking the Fourier
transform of Eq.~(\ref{noisespec}) leads to
\beq
\langle \tilde{\eta_c}(\vec{k},\omega)\tilde{\eta}(\vec{k'},
\omega') \rangle_{\grave{\eta}}
= \frac{2D(k,\omega)}{(2\pi)^{d}}\delta^d(\vec{k} + \vec{k'})
\delta(\omega + \omega')\,\, ,
\eeq
where $D(k,\omega) = D'|\vec{k}|^{-2\rho}|\omega|^{-2\varrho}$, and $D'$ is
a function of $\rho$ and $\varrho$. The analysis of the power law behavior
for both EW and MWV cases we get the following relations between the new
roughening exponents (primed) and the old ones:
\begin{eqnarray}
\beta_1' = \beta_1 + \frac{1}{2}\rho + \varrho & ; & \chi_1' = \chi_1
+ \rho + 2\varrho \,\, . \\
\beta_2' = \beta_2 + \frac{1}{4}\rho + \varrho & ; & \chi_2' = \chi_2
+ \rho + 4\varrho\,\, .
\end{eqnarray}
Clearly, $z'_1\equiv \chi_1'/\beta_1' = 2$ and $z'_2\equiv
\chi_2'/\beta_2' = 4$ are satisfied because $\chi_1/\beta_1 = 2$ and
$\chi_2/\beta_2 = 4$. These results agree with previous treatments
\cite{Lam91,thh,Med89}.
\section{Summary and Conclusions}
\label{summary and conclusions}
In this work, we have presented a detailed quantitative
analysis of a combined linear stochastic growth equation,
which incorporates both EW and MWV type of behavior.
In particular, we have identified the relevant scaling
variables and calculated in detail the behavior of the
surface width, and various correlation functions in
different regimes. Our analysis shows that the CLG model
possesses MWV type of
scaling regime at early times ($t \ll t_c$) and crosses
over to EW type of growth for later times. In a finite system it is,
however, possible that the system maintains MWV type of growth dynamics
for all times
given that the prefactor of the fourth order gradient term is very
large ($L \ll \ell^\ast$).
Additionally, the equal time and saturated correlation
funtions reveal interesting and complicated crossover behavior
between the two limits.
We should also note that in the appropriate limits,
our results completely agree with the previously derived
results for
concerning the surface width and the equal time correlation functions of
the EW and MWV equations \cite{Edw82,Lam91,Nat92,Ama93,Sar94,thh}.
We have also recalculated the scaling
exponents in Sec. \ref{applications}
for coloured noise that has long range spatial and temporal correlations.
Our results show that the long--range
part of the noise changes the exponents, in analogy to the
case of the nonlinear KPZ equation \cite{Med89}.

The solvability of the CLG model allows us to address
other important problems related to surface growth problems
as well.
A particularly interesting case concerns the influence of
rough initial conditions.
In Sec. \ref{applications} we demonstrate
that the effect of initial roughness in $d=2$ disappears as a power
law in time $\sim t_0/t$ if the surface has initially
roughened through the CLG growth dynamics for a time $t_0$. In general,
the initial roughness vanishes as $(\xi_s/\xi)^d \sim t^{-d/z}$ for EW and MWV
models as $\xi \gg \xi_s$. For a general growth process the behavior of this
transient is easy to evaluate by Fourier transforming the exponentially
damped initial height--height correlation function (Eq.~(\ref{furu})).

Finally, to make the connection to experiments more concrete
we have proposed the use of the parametrized correlation
functions obtained for the CLG model for
fitting of diffusive x--ray refletivity data.
In Sec. \ref{applications}
we establish a relation between the parameters $\nu_1$ and $\nu_2$ of the
CLG growth model and the experimental fitting parameter $\xi$ (cf. Eq.
(\ref{xiku})). Moreover,
both the small ($x \ll \xi$) and the large ($x \gg \xi$) scale behavior
of the correlation function of the CLG model are found to be different from
the experimentally used fitting functions. This also implies that the measured
decay exponent $\chi$ from heuristic fitting functions bear no direct relation
to the roughness exponent of the CLG model. We hope that these calculations
will be useful in future experiments on surface growth. \\

Acknowledgements: J. K. wishes to thank H. T. Dobbs, M. Funke and F. K\"onig
for stimulating discussions. This work has in part been supported by a joint
grant between the Academy of Finland, and Deutscher Akademischer
Austauschdienst.

\bigskip\noindent
E--mail addresses: {\tt majaniem@fltxa.helsinki.fi},
{\tt ala@fltxa.helsinki.fi}, {\tt j.krug@kfa-juelich.de}
\pagebreak
\section*{Appendix A: Substrate Roughness for $z_s = z$}

Here we explicitly evaluate the integral (\ref{wi2}) in the special case
$z_s = z$. Taking the derivative of (\ref{wi2}) with respect to $\tau$ and
reintegrating we obtain
\begin{equation}
\label{wisimple}
w_i^2 = \frac{\Omega_{d-1} A_0 \Gamma(d/z)}{z - d} \xi_s^{z-d} [ (\tau + 1)^{1
- d/z} - \tau^{1-d/z} ].
\end{equation}
Three cases have to be distinguished.

{\em (i)} $z > d$: In this case both the substrate and the growing surface
are rough with roughness exponent $\chi = (z-d)/2 > 0$. The expression
(\ref{wisimple}) then has a finite limit
$w_i^2(0) = \Omega_{d-1} A_0
\Gamma(d/z) \xi_s^{z-d}/(z-d)$ for $\tau \to 0$. For $\tau \ll 1$ the
roughness remains essentially unchanged,
\beq
\label{wisimpleshort}
w_i(\tau)^2 \approx w_i(0)^2 (1 - \tau^{1-d/z}),
\eeq	
while for $\tau \gg 1$ it decays as
\begin{equation}
\label{wilongtime1}
w_i^2(\tau) \approx (1 - d/z) w_i^2(0)\tau^{-d/z}
\eeq
in accordance with (\ref{wi0}).

{\em (ii)} $z = d$: Substrate and surface are logarithmically rough.
Taking the limit $(z - d) \to 0$ in (\ref{wisimple}) gives
\begin{equation}
\label{wilog}
w_i^2(\tau) = (\Omega_{d-1} A_0/z) \ln(1+1/\tau)
\end{equation}
which diverges for $\tau \to 0$. Clearly (\ref{wilog}) is valid only for
$\xi > k_{\max}^{-1}$, i.e. $\tau > \tau_0 = (k_{\max} \xi_s)^{-z}$; the
initial
roughness is $w_i(\tau_0)^2 \approx \Omega_{d-1} A_0 \ln (k_{\max} \xi_s)$.
For $\tau_0 \ll \tau \ll 1$ the substrate contribution decays slowly, as
$w_i(\tau)^2 \approx \Omega_{d-1} A_0 \ln(1/\tau)$, while for $\tau \gg 1$
(\ref{wilog})
reduces to
\begin{equation}
\label{wiloglong}
w_i^2(\tau) \approx (\Omega_{d-1} A_0 /z) \tau^{-1}
\eeq
which is of the form (\ref{wi0}) with a prefactor of order unity.

{\em (iii)} $z < d$: Substrate and surface are smooth, and the width is
microscopic at all times. In this case (\ref{wisimple}) combines two
power law decays: For $\tau \ll 1$
\begin{equation}
\label{wismooth1}
w_i^2(\tau) \sim \xi_s^{-(d-z)} \tau^{-(d/z-1)} \sim \xi^{-(d-z)}
\end{equation}
independent of $\xi_s$, while for $\tau \gg 1$ one finds
\begin{equation}
\label{wismooth2}
w_i^2(\tau) \approx [\Omega_{d-1} A_0 \Gamma(d/z)/z] \xi_s^{-(d-z)} \tau^{
-d/z} \sim \xi_s^{-(d-z)}(\xi_s/\xi)^d
\end{equation}
as in the previous two cases.
\pagebreak

\section*{Appendix B: Derivation of the Exponent $\delta$}

Here we present a simple way of computing the exponents $\delta$ and $\gamma$
characterizing the decay of the height correlation scaling function
(\ref{genscal}) for the generalized linear equation (\ref{genlinear}) in
one dimension. It follows immediately from (\ref{genlinear}) that the
Fourier transform $\hat F(q)$ of $F(s)$ is, up to some factor,
\beq
\label{Fhat}
\hat F(q) \approx q^{-z} (1-e^{-q^z}).
\eeq
We now exploit the fact that by definition, $\hat F(q)$ can also be written as
\cite{harvey}
\beq
\label{Fhatexpand}
\hat F(q) \approx \sum_{n=0}^\infty \frac{A_n}{n !} (-\imath q)^n
\eeq
where
\beq
\label{moments}
A_n = \int_{-\infty}^\infty ds \; s^n F(s)
\eeq
is the $n^{\rm th}$ moment of the real space scaling function.
Comparing (\ref{Fhat})
to the power series (\ref{Fhatexpand}), we obtain the relation
\beq
\label{momentrelation}
\vert A_{zk} \vert \approx \frac{1}{k+1} \frac{(zk)!}{k !}, \;\;\; k=0,1,2....
\eeq
The asymptotics of $F(s)$ can then be extracted by comparing the behavior of
(\ref{momentrelation}) for large $k$ with that obtained from the ansatz
(\ref{genscal}) inserted into (\ref{moments}); assuming the amplitude $c$
in (\ref{genscal})  to be real, the integral (\ref{moments}) is easily
evaluated in the saddle point approximation. The leading behavior is
$A_n \sim n^{n/\delta}$ which, compared to $\vert A_{zk} \vert
\sim k^{(z-1)k}$ from (\ref{momentrelation}), implies that
$\delta = z/(z-1)$. Moreover, from the integral (\ref{moments}) one obtains
a power law factor $\sim n^{1/\delta - 1/2 - \gamma/\delta}$ which has to
be matched to the factor $(k+1)^{-1}$ that appears in (\ref{momentrelation});
this forces $\gamma$ to take the value $\gamma = 1 + \delta/2$. Both results
agree with the detailed calculations of Sec. \ref{Correlation}.

\newpage
%
%\appendix
%
%\section{}
%\label{}
%
%
%
%

%
%
%
%
\newpage
\pagestyle{plain}
{\noindent\Large\bf Figure Captions} \\
\bigskip\noindent
\begin{itemize}
\item [{Fig. $1$}] A log--log plot of the surface width
$w^2(t)$ of the CLG equation for several values
$|\nu_2|$ and $\nu_1$ in $d=1$, with $L = 10 000$.
The two scaling regimes of the surface width $w^2 \propto t^{3/4}$
for $t \ll t_c$ and $w^2 \propto t^{1/2}$ for $t \gg t_c$ are
clearly visible.
For the topmost curve $\nu_1=1$ and $|\nu_2| = 10^{-8}$ and the
crossover time $\ln t_c \equiv \ln (|\nu_2|/\nu_1^2)\approx -18$ is
shown. For the second curve $\nu_1=|\nu_2| = 1$, and
$\ln t_c = 0$. Finally, the two overlapping lowest curves
represent the CLG surface width with $\nu_1=1$ and $|\nu_2| =
10^{8}$, and the pure MWV case with $\nu_1 = 0$, $|\nu_2| =
10^{8}$. In both cases the saturation time
$\ln t_s\approx 18$.
\item[{Fig. $2$}]
Decay of the substrate contribution to the surface width for dimensionality
$d=2$ and dynamic exponents $z_s = z = 4$ (full upper curve) and
$z_s = z = 2$ (full lower curve). The full curves show the exact expressions
derived in Appendix B, while the dotted curve shows the interpolation formula
(\ref{interpolation}). The dashed lines indicate the asymptotic power laws.

\item[{Fig. $3$}]
(a) The correlation length $\xi$ as a function of the two dynamical 
correlation
lengths $\xi_1 \equiv (2\nu_1 t)^{1/2}$ and $\xi_2 \equiv (2|\nu_2| t)^{1/4}$
of the CLG model.
(b) The relative error $E \equiv (\xi_p - \xi)/\xi$ between the plane fit
$\xi_p = \alpha_1\xi_1 + \alpha_2\xi_2$ and $\xi$ (Eq. (\ref{geeku}).
The {\em average} relative error is about $1 \%$. The maximum error in the
figure ($10 \%$) occurs for large $\xi_1$ and small $\xi_2$.
\end{itemize}
\end{document}